# Switching of Perpendicularly Polarized Nanomagnets with Spin Orbit Torque without an External Magnetic Field by Engineering a Tilted Anisotropy

Long You[1], OukJae Lee[1], Debanjan Bhowmik[1], Dominic Labanowski[1], Jeongmin Hong[1], Jeffrey Bokor[1], Sayeef Salahuddin[1,2]

[1] Department of Electrical Engineering and Computer Sciences, University of California at Berkeley, Berkeley, California 94720, USA,

[2] Materials Sciences Division, Lawrence Berkeley National Laboratory, Berkeley, California 94720, USA

Spin orbit torque (SOT) provides an efficient way to significantly reduce the current required for switching nanomagnets. However, SOT generated by an in-plane current cannot deterministically switch a perpendicularly polarized magnet due to symmetry reasons. On the other hand, perpendicularly polarized magnets are preferred over in-plane magnets for high-density data storage applications due to their significantly larger thermal stability in ultra-scaled dimensions. Here we show that it is possible switch a perpendicularly polarized magnet by SOT without needing an external magnetic field. This is accomplished by engineering an anisotropy in the magnets such that the magnetic easy axis slightly tilts away from the direction, normal to the film plane. Such a tilted anisotropy breaks the symmetry of the problem and makes it possible to switch the magnet deterministically. Using a simple Ta/CoFeB/MgO/Ta heterostructure, we demonstrate reversible switching of the magnetization by reversing the polarity of the applied current. This demonstration presents a new approach for controlling nanomagnets with spin orbit torque.

Spin orbit coupling (SOC) and/or broken inversion symmetry in vertical heterostructures can generate accumulation of spins when a charge current is flowing through them. In doing so, it can exert a torque on an adjacent magnet [1-8]. Indeed, high Z metals (Ta, Pt, W, etc.) with strong SOC have been used to inject spin currents into adjacent ferromagnetic layers and thereby to induce magnetic switching, oscillation, domain wall movement etc [1-5,7-9]. In a typical heterostructure such as Ta/CoFeB/MgO (from the bottom), an in-plane current flowing in the x-direction (electrons flowing in the –x direction) generates $\hat{\sigma} = \hat{y}$ polarized spins that accumulate at the Ta/CoFeB interface. Therefore, if the ferromagnet (CoFeB) is polarized in-plane, the spin accumulation can rotate it to the +y direction by a Slonczewski-like torque $\vec{\tau}_{sl} = \tau_{sl}^0 \ (\hat{m} \times \hat{\sigma} \times \hat{m})$ [10]. The magnet can be switched to –y direction by reversing the polarity of the current. Thus a deterministic switching is possible by an in-plane current when the magnet is also polarized in-plane. However, if the magnet has a perpendicular magnetic anisotropy (PMA), an in-plane current and resultant in-plane spin accumulation cannot break the reversal symmetry. Consequently, no deterministic switching of the PMA magnet can be obtained. Because of this reason, an external magnetic field has to be applied in-plane in the same (or opposite) direction of the current flow which breaks the symmetry and makes it possible to switch a PMA magnet with an in-plane current [11]. It is, however, desirable to switch the magnetization without needing an external magnetic field. At the same time, PMA is more suitable for scaling magnets to ultra small dimensions while retaining reasonable thermal stability [12-14]. This means that alternate ways need to be found that can lead to symmetry breaking and make it possible to switch perpendicularly polarized magnets with in-plane current without having to apply an external magnetic field. In this paper we report on such a scheme where we have fabricated magnetic nanodots from a heterostructure stack of Ta/CoFeB/MgO such that their easy axes are slightly tilted from the film normal. This tilting breaks the symmetry with respect to an in-plane SOT. Indeed, pulsed currents flowing in-plane deterministically switch the magnetization up and down without any external magnetic fields.

The essential idea is described in Fig. 1. Imagine a magnet that has an easy axis on the x-z plane but slightly tilted away from the z axis to x-axis as show in Fig. 1(a). Here x-y is the film plane. The actual device is made of a heterostructure consisting of Ta/CoFeB/MgO as shown in Fig. 1(b). For brevity, when the magnet is in the (x,z) quadrature, we shall call it the 'up' position.

Similarly when it is in the (-x,-z) quadrature, we shall call it the 'down' position. Now imagine a current is applied in the x direction such that a y polarized spin accumulation is generated at the Ta/CoFeB interface. Since, y is orthogonal to the x-z plane, regardless of whether or not there is a tilt in the easy axis of the magnet in the x-z plane, no symmetry is broken and therefore the magnetization cannot be switched deterministically. However, this situation changes if a current is applied in the y or -y direction. This will generate a spin accumulation polarized in the –x or +x direction respectively (for Ta). If the magnetization is in the x-z plane and tilted from the z-axis, the direction of spin polarization (x or -x) is no longer symmetric with respect to the tilted easy axis of the magnet and this symmetry breaking can lead to deterministic reversal of the magnetization [15]. For example, let us consider the situation when the magnet is polarized up and a current is flowing in the y direction leading to a spin polarization in the –x direction. If this accumulation is strong enough to rotate the magnetization to the –x direction, in the process, the magnet would cross over the hard axis, shown by the dotted line in Fig. 1(a). This means that when the current is turned off, it is preferable for the magnet to now go to the 'down' position in the (-x,-z) quadrature. On the other hand, if a current is applied in the –y direction, the spin accumulation is +x polarized. This means starting from up, the magnet, in the process of rotating to +x direction, does not cross over the hard axis. And so when the current is turned off, it is preferable for the magnet to go back to the up state. Thus, a current along +y switches the magnet from up to down. Similarly, a current along –y will switch the magnet from down to up. When the magnet is up , a current along –y does not change the state of the magnet. Similarly, when the magnet is down, a current along +y cannot change the state of the magnet. Thus a fully deterministic switching of the perpendicular magnetization can be achieved without an external magnetic field.

Our fabricated devices are based on a stack consisting of Ta (10 nm)/CoFeB (1 nm)/MgO (1 nm)/Ta (~ 3nm) (from the bottom). Vibrating Sample Magnetometer (VSM) measurement shows excellent perpendicular anisotropy in the thin films (SI section 1). First small nanomagnets are patterned with a high aspect ratio (80 nm x300 nm). Next, a wedge shape was created by carefully controlling the thickness of the hard mask and Ar$^+$ ion milling such that the thickness of the magnet gradually goes to zero on one side(details are described in Methods and Fig. S2). Fig.2a shows an SEM micrograph of a typical nanodot and the underneath Hall structure for transport measurement. Fig. 2b shows an atomic force microscopy (AFM) image of the nanodot. A wedge like feature is clearly seen. The depth profile shown in Fig. 2.c indicates that MgO with lateral size of around 50 nm in the left part of nanomagnet was completely etched, and CoFeB wedge was created. Because MgO was etched out, the magnetization in the wedge region no longer has a perpendicular anisotropy and will follow the edge of the wedge to minimize magnetostatic enerrgy, thereby providing an overall tilt to the perpendicular magnet, directed away from the edge. Micromagnetic simulations (see SI section 2) confirm that this indeed is the case, as we will discuss later.

To experimentally demonstrated that the tilt, we have performed anisotropic magnetoresistance (AMR) measurements to investigate potential tilting of the easy axis. Note that all the experiments in this work were done at ambient temperature. In the coordinate axes as shown in Fig. 3(a) and (b), the dependence of AMR resistivity on the in-plane magnetic field ($H_{inp}$) rotation angle $\theta$ can be written as

$$\rho(\theta) = \rho_\perp + (\rho_{//} - \rho_\perp)\cos^2\varphi(\theta) = \rho_\perp + \Delta\rho\cos^2\varphi(\theta) \qquad (1)$$

where $\varphi$ is the angle between the magnetization $M$ and the direction of current and $\rho_\perp$ ($\rho_{//}$) is the resistivity when the magnetization is perpendicular (parallel) to the current direction. Let us consider the case when current is flowing in the -x direction and a $H_{inp}$ of a specific amplitude is rotated in-plane, starting from the –x axis and ending on the +x axis. Since the current is applied along the x axis, from Eq. (1), the AMR is the maximum when the $H_{inp}$ is also along the ±x direction. AMR is lower when $H_{inp}$ is rotated away from x-axis. Notably, if the magnetic easy axis is purely out-of-plane to begin with, applying the same magnitude of $H_{inp}$ in the +x or –x direction should give the same AMR value. By contrast, if the magnet is tilted on the +x-axis, one would expect to see a larger AMR when the $H_{inp}$ is in the +x-direction in comparison to when the field is in the –x direction. This is exactly what we observe in our experiments. Starting from an initially up position, the AMR is larger when $H_{inp}$ is in the +x direction compared to when it is in the direction and the AMR goes down away from the x-axis (see Fig. 3c). On the other hand, starting from an initial down position, the AMR is larger when $H_{inp}$ is in the -x direction (see Fig. 3d). This asymmetric AMR curve is also observed when magnetic fields of other magnitudes are applied such as 500 Oe, 800 Oe, 2000 Oe etc. (See SI section 3 for details). These data show that the easy-axis of the magnetization is slightly tilted from the z-axis and lies along (x,z) quadrature to (-x,-z) quadrature. The observed behavior can be modeled well from equation (1) including the fact that the lowest value of AMR arises slightly away from the y-axis and the sign of the shift depends on the starting polarization of the magnet, i.e. 'up' or 'down'. Finally, we have done AMR measurements with large $H_{inp}$ (~3000 Oe). Fig. 3(e,f) show AMR data at this field starting from up and down positions respectively. No asymmetry is observed between the AMR recorded for $H_{inp}$ applied in the +x and −x direction. This is expected because now $H_{inp}$ is strong enough to overcome the tilted anisotropy and the magnet follows the magnetic field. Thus, from the symmetric AMR along the x-axis at high field and asymmetric AMR at low field, the existence of a tilted easy axis in the nanodots is confirmed. From the sign of the asymmetry, when the CoFeB wedge is in −x as shown in Fig. 2 (c), one can also conclude that the easy axis lies on the x-z plane, slightly tilted towards x from z, as shown in Fig. 1(a). From AMR measurement, we can also estimate the angle of the tilt which comes to be ~2° with an error of 1° due to the uncertainty in the angle of the in-plane field. At the same time, from comparing the slant in the R-H loop of an anomalous Hall measurement for the wedged dot to a circular dot, another estimate for the tilt amplitude can be made. This measurement gives the tilt angle to be ~5° (see SI section 5). Based on these measurements, we estimate the tilt angle to be between 2-5°.

With such a tilted easy axis of the nanodots, the switching scenario described in the preceding paragraphs, namely a deterministic switching from up to down with a current in the y direction and from down to up with a current in the −y direction should be possible. To demonstrate that, we have performed extensive AHE measurements. The corresponding hysteresis loop is shown in Fig. 4(a). Fig. 4(b) shows switching data at room temperature in the presence of zero external field. The magnet is saturated to down first by applying a strong vertical magnetic field in −z direction. Next, the field is turned off and a current pulse (~1 sec duration) of 1.25 mA ($2.5 \times 10^7$ A/cm$^2$) is applied in the −y direction. After the pulse is off, an AHE measurement is done by applying a small sense current. The AHE resistance as shown in the bottom panel of Fig. 4(b) shows that the magnet is now up. Next, a current pulse of the same amplitude is applied in the +y direction. A subsequent AHE measurement finds the magnet in the down direction. In the process, value of $R_{AHE}$ switches by ~0.24 Ω. Comparing with Fig. 4(a), this shows that a full switching of the magnet takes place. The full R-I loop measured in this way is provided in SI section S.6.1.

We tested the reproducibility of the switching process by repeatedly reinitializing and subsequently switching the nanodot in one particular direction 10 times. Fig. 4(c) shows the observed behavior. Fist the magnet is polarized in the up direction. Next a $+I_y$ (current flowing in the +y direction) is applied and a subsequent AHE measurement is done to find if the magnet has switched. Following that, the magnet is polarized up again with a magnetic field and again a $+I_y$ is applied. This process is repeated 10 times. If the switching happens all 10 times the switching probability ($p_{sw}$) is 1. Fig. 4(c) shows the switching probability for 4 different devices. We see that a 100% switching is observed for all devices. It is to be noted that starting from up $-I_y$ is not supposed to change the state of the magnetization. To test that, the same experiment as described above is performed with the only exception that the current pulse is now applied in the –y direction. Among 4 different devices, each going through 10 repeated trials, only device 3 shows a small probability (0.2) of switching. We attribute this to thermal heating of the dots due to repeated pulsing. Next a similar experiment is done on the same 4 devices but now starting from an initial down position and applying a $-I_y$. The data is shown in Fig. 4(d). Again an almost perfect switching probability is observed. Application of a $+I_y$ is not expected to induce any switching in this case. We found that that is largely the case with some small errors in device 1(0.3) and device 3(0.1). This is again attributed to increased heating due to repeated pulsing. Fig. 4(b), (c) and (d) demonstrate deterministic switching of the magnetization without any external magnetic field by a current flowing along the y-axis as described in the beginning of this paper. As it was mentioned, a current flowing in the x direction cannot break the reversal symmetry for our magnet which lies on the x-z plane. Indeed, applying current pulses in the x direction did not lead to any deterministic switching of the magnet (see SI section S.6.2). A final test was done by using 200 nm x 200 nm magnets on a symmetric Hall bar without any wedge fabricated from exactly the same stack. Applying pulse current to such dots without any external magnetic field also did not lead to any deterministic switching (see SI section S.6.2).

Since there is a structural symmetry breaking due to the wedge shape in the fabricated nanodots in the x-direction, it is possible to get a Rashba Field in the out-of-plane direction when a current is flowing in the y direction [20]. In principle, such a Rashba Field can break symmetry and contribute to the switching of the nanodot. Second harmonic measurements in our devices show similar values for longitudinal and transverse field components to that have been reported for similar structures in literature [6,20](see SI section 8). To test the existence of an out-of-plane field in our devices, we have measured the threshold current needed to switch the magnetization in presence of an out-of-plane magnetic field. The basic idea is that if a current induced Rashba field is present in the out-of-plane direction, the *M* vs. $H_z$ hysteresis loop will be shifted left or right from its symmetric-around-zero position as shown in Fig. 4(a). Fig. 4(e) shows the variation of threshold current flowing in the +y direction ($I_y^{th}$) as a function of the absolute value of an applied field in the z direction. First, note that at zero current the $H_c$ and -$H_c$ have the same absolute value, which is 175 Oe (note that this is a different device than that shown in Fig. 4(a)). If a Rashba field is present, a gap between $H_z$ and $-H_z$ should open up as the amplitude of applied current goes up. Therefore, the most important region to look at will be where the current amplitude is close to the threshold amplitude needed for zero magnetic field switching. However, in Fig. 4(e) we observe that at these high current levels there is hardly any difference between $H_z$ and $-H_z$. A similar trend is seen when a current is applied in the $-y$ direction in Fig. 4(f). This indicates that the contribution of an out-of-plane Rashba field, if any, is minimal in our devices.

In the intermediate range where neither the applied field nor the current alone is strong enough to switch the magnet, a gap between $H_z$ and $-H_z$ can be observed. Note that in this intermediate range there is significant difficulty in ascertaining the exact value of the threshold current. Essentially, after setting a specific absolute value of the magnetic field, say 75 Oe, multiple experiments had to be done by varying the amplitude of the applied current pulse to find the threshold value. These repeated trials lead to some uncertainty as to the exact value of the threshold current and also heating of the samples which can also influence threshold current non-trivially. We note that for a given absolute value of the magnetic field the difference between the switching currents never exceeded 10% of the applied pulse amplitude (see Fig. 4(e) and 4(f)). This is well within the experimental error in this intermediate region. As either the current or the field amplitude increases, the switching becomes easier and the uncertainty in the switching current goes down. In addition, if one tries to associate a Rashba field for this intermediate region, for $+I_y$ it will be pointing in the $-z$ direction from Fig. 4(e). But for $-I_y$, it is still in the $-z$ direction (see Fig. 4(f)), although reversing the current polarity should have reversed it. Because of the aforementioned two reasons, we believe that the gap observed between the magnetic field values in the intermediate region is an experimental artifact.

Detailed micromagnetic simulations were performed to elucidate on the underlying physics (see SI section 2 for details). A mesh was set up to capture the wedge in the magnet. With a magnet that closely follows the experimental dimensions (total 300 nm in length with 60 nm wedge, 80 nm wide and 1.2 nm thick in the thickest area), the first set of tests were aimed at finding the equilibrium magnetization of the magnet. Starting from an initial condition of mostly up (with a slight component in the y-direction to initiate the dynamics), when the magnet was allowed to relax, it went to an up position with a small component along the x direction, which is away from the wedge (see Fig. 5a), exactly as it is observed in the experiments. The opposite happens when one starts from an initial condition of mostly down (see Fig. 5b). This effect is understandable from magnetostatics. The magnetization in the wedge area has no perpendicular anisotropy because the MgO has been completely etched off. As a result they follow the edge of the wedge such that the surface magnetic charge is minimized. Furthermore, they also try to align with the spins in the perpendicular region due to exchange coupling. As a result, they point in the +x direction when bulk of the magnet is +z and vice versa. What is reassuring is the fact that, the overall tilt comes to be around 3° (see SI section S.2.2), which is very similar to what we have measured experimentally. Now starting from an initial magnetization along the x-axis, as it will be when a current is applied along the y-axis, depending on whether it started from +x or −x, the magnetization goes to +z or −z direction after the current is turned off, exactly as we have seen in our experiments (Fig. 5c and 5d). Therefore, the micromagnetics can completely reproduce our experimental observations. It also shows that the switching process is likely to be mediated by domains that form at the intersection of the wedged and non-wedged regions.

To summarize, we have demonstrated that, by engineering a slightly tilted anisotropy axis in a perpendicular magnetic nanodot, it is possible to reversibly and deterministically switch it without applying any external magnetic field. The tilted easy-axis breaks the symmetry that is otherwise present between an in-plane current induced spin accumulation and a PMA magnet and makes it possible to switch it without needing to break the symmetry with an external magnetic field. Experimentally measured tilt in the magnetic easy axis direction is consistent with the polarity of the current needed to switch the magnet from up to down and vice versa, as it will be predicted by a Slonczewski-like torque. Although, we have used a sophisticated fabrication method to engineer a tilted anisotropy in otherwise perpendicular magnetic stack, it is conceivable that a precise control of the tilt angle could be possible at the film level by combining a PMA hard layer (such as Pt/Co) with an in-plane soft layer (IMA, such as CoFeB, NiFe, Ni). Our demonstration of reversible switching without an external magnetic field could lead to combining the promise of ultra low current requirements in spin Hall systems to the scaling advantages of perpendicular media for next generation storage and other spintronic applications.

## Methods

### Experiments

A stack comprising a thermally oxidized Si substrate/Ta (10 nm)/CoFeB(1 nm)/MgO (1 nm)/Ta (~3 nm) was fabricated into Hall bars by photolithography and argon ion milling. The Hall bars contained the entire thin-film stack with the region outside the Hall bars etched down to the insulating oxidized Si substrate. We used electron beam lithography with ZEP520A resist and ion milling to define the current channel and the detection channel, both being 500 nm long. The key challenge is to form the desired device pattern in the Ti layer, which acts as an etching mask and is used as a pattern transfer. First, 170 nm thick poly methyl methacrylate (PMMA) was patterned by a second stage of aligned electron beam lithography onto a set of rectangle shapes (with aspect ratios varying from 4 to 7). Then 4 nm of Ti was deposited normal to the substrate. Ti formed a wedge shape at the two sides along the long axis of the nanomagnet due to shadowing by the resist mask during evaporation of the Ti film [23], as shown in Fig. S2 a. This was followed by obliquely depositing 3 nm at the tilting angle of $\theta$ (15° - 45°) from the substrate normal, shown in Fig. S2 b. This prevents Ti deposition on one side ascribing to the shadowing effect of resist. After the lift-off process, rectangular Ti nanodots with non-uniform thickness were formed, as shown in Fig. S2 c. In particular, one side has wedge shaped Ti with maximum thickness of 3nm, while the thickness of other part is around 7 nm. Argon ion milling was used to etch the stack in the region outside the dot patterns down to the bottom tantalum layer.

Meanwhile, a wedge shape was created at one side along the length (long axis) of the nanomagnet, as shown in Fig.1 b and Fig. S2 d.

For the anomalous Hall resistance measurement, current was applied using a current source and the Hall voltage was measured using a nanovoltmeter. The same current source was used to apply current pulses for switching and switching phase diagram measurements. A unipolar current pulse with a duration of ~1s was used for all measurements. Anisotropic magnetoresistance was measured using the standard four-probe technique. All measurements were performed at room temperature.

**Acknowledgement:**

This work was supported in part by the DOE Office of BES, NSF E$^3$S center, STARNET FAME center. DB acknowledges support from an Intel Fellowship.

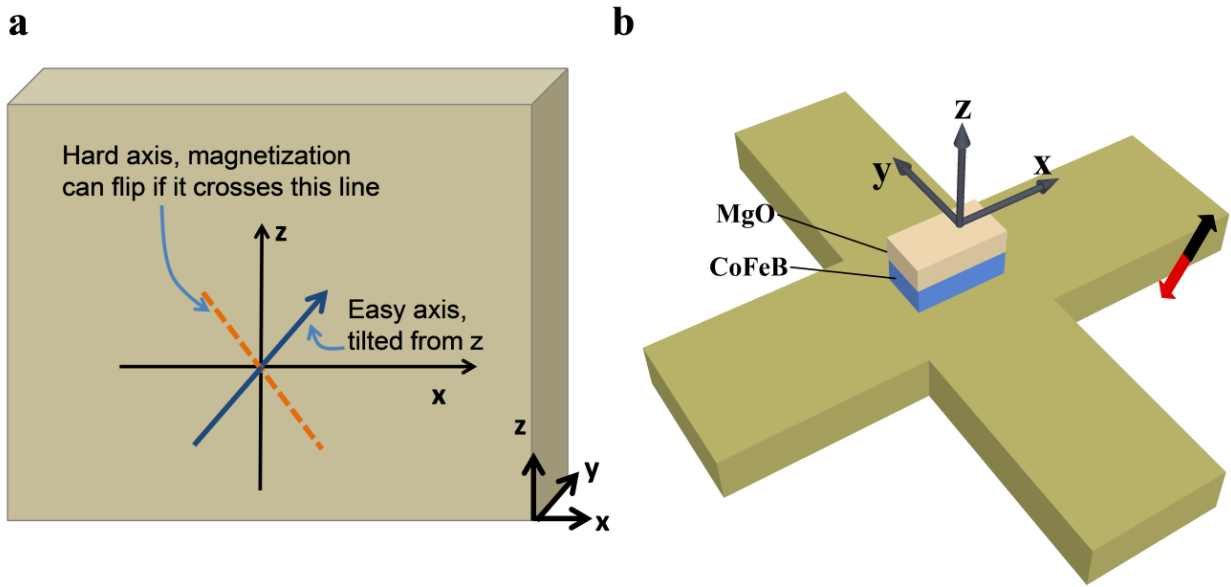

**Figure 1: Schematic description of switching mechanism and device structure.** (a) Orientation of the easy axis and hard axis of an otherwise perpendicularly polarized magnet with slightly tilted anisotropy. (b) The fabricated device structure. The underneath Hall bar is composed of Ta and is used to generate spin accumulation through spin-orbit interaction and also to detect the direction of magnetization through anomalous Hall effect measurement. Current flowing along the short direction of the magnet (along the y-axis) leads to a deterministic switching.

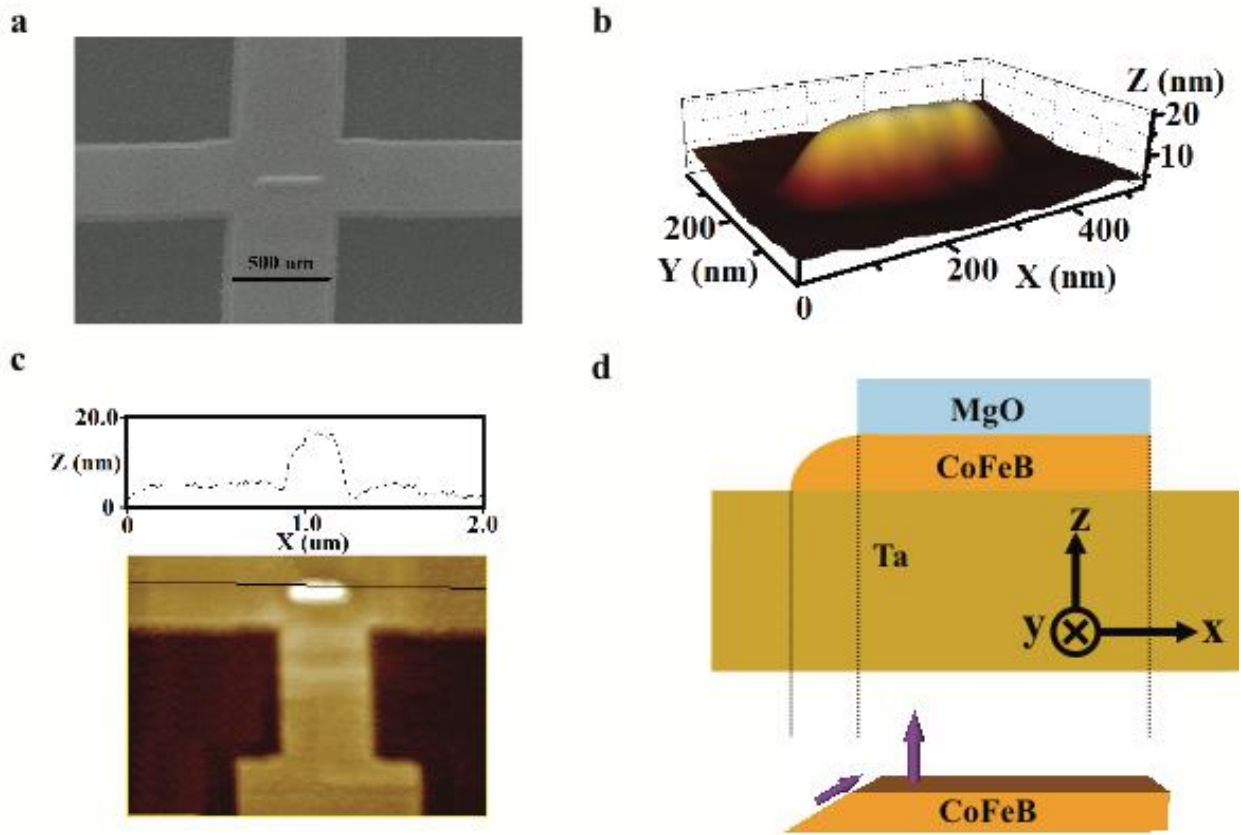

**Figure 2: Engineering a tilted easy axis.** (a) Scanning electron micrograph of the fabricated device (b) three-dimensional morphological characterization of nanomagnet by atomic force microscopy showing a wedge like structure. (c) Quantitative measurement of the wedge using atomic force microscopy. (d) Schematic showing the structure and the tilt coming from the spins aligning to the wedge

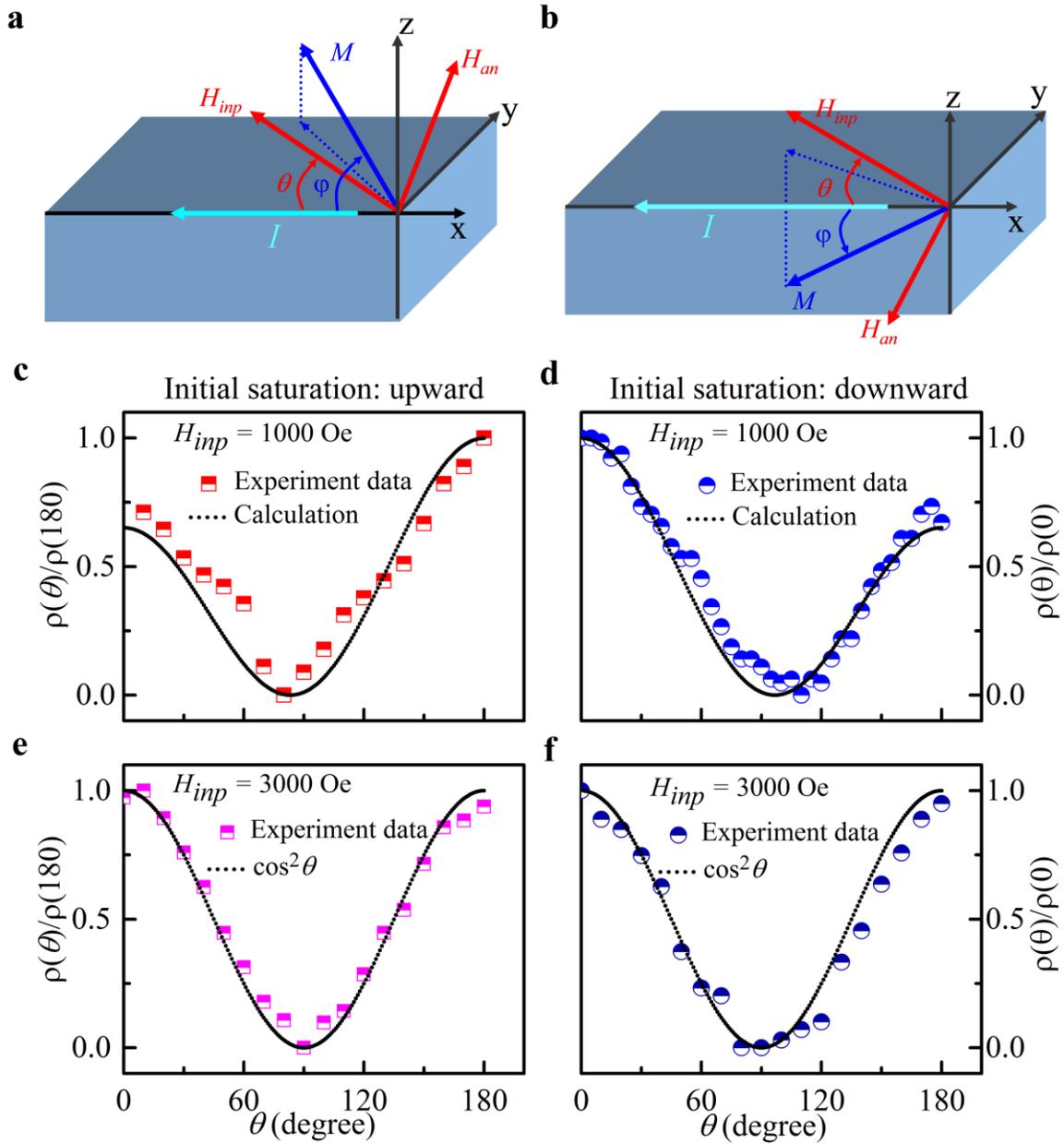

**Figure 3: Estimation of the tilt direction.** The schematic configuration of the coordinate system used in the analysis of the AMR data. (a) and (b) show the configurations for the initial polarization of up and down respectively. The sense current for AHE measurement flows along $-\hat{x}$ direction pointing to the long axis of the nanomagnet. $\theta$ is the angle between the directions of the measurement current and the external in-plane field $H_{\text{inp}}$. $\theta = 0$ corresponds to the orientation where the applied field is aligned to the direction of current flow ($-\hat{x}$). $\theta$ was taken

as positive for a anticlockwise rotation of the field from current. $\varphi$ is the angle between *M* and *I*. Calculated and measured values of AMR resistivity are shown when $|H_{inp}|$=1000 Oe and the magnet was initially polarized (c) upward and (d) downward. Calculated and measured values of AMR resistivity are shown when $|H_{inp}|$=3000 Oe and the magnet was initially polarized (e) upward and (f) downward.

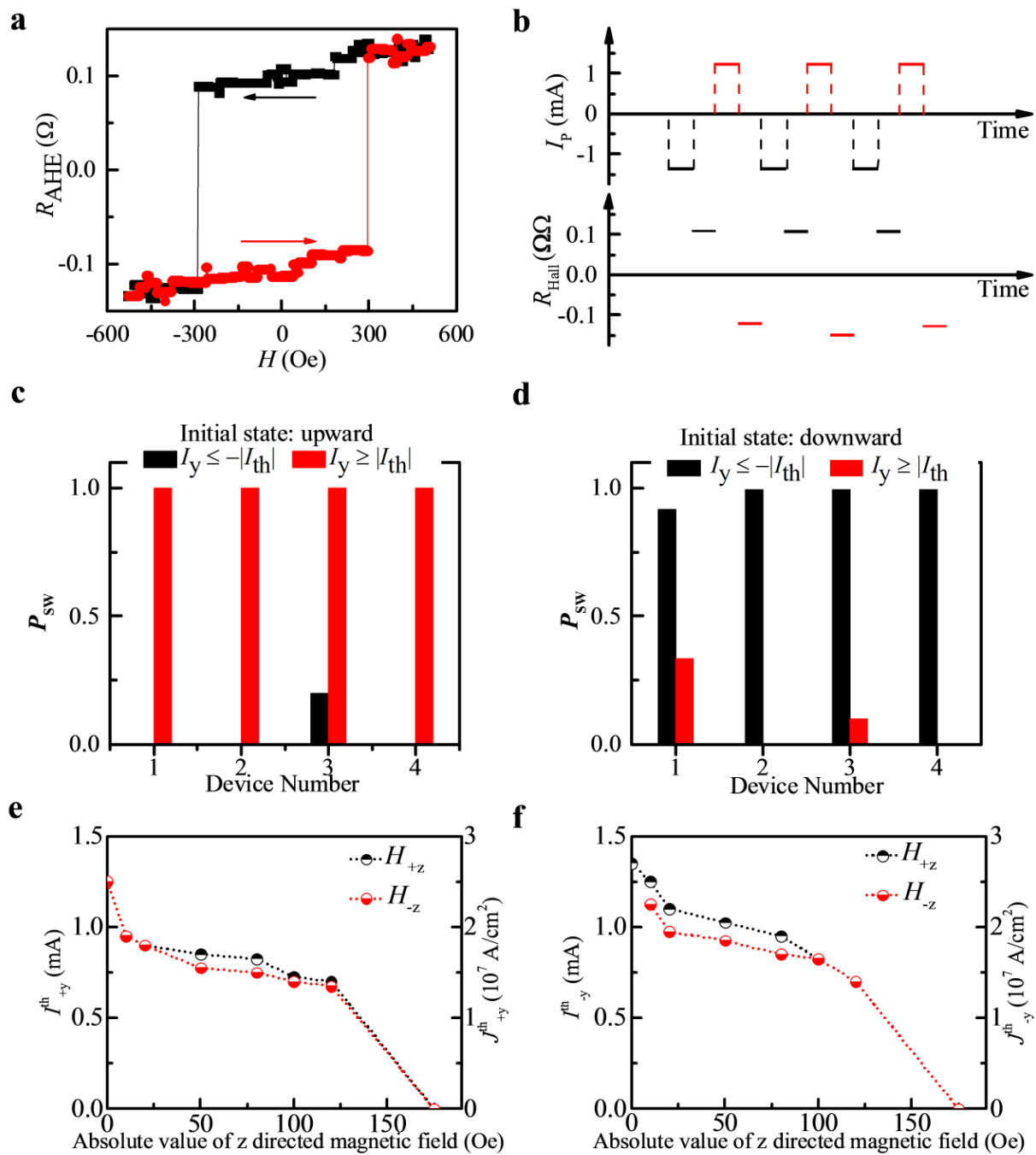

**Figure 4: In plane current induced switching without external magnetic field.** (a) M-H hysteresis characteristic measured by AHE. Positive $R_{AHE}$ corresponds to "upward" magnetization, while negative $R_{AHE}$ corresponds to "downward" magnetization. (b). Evolution of $R_{AHE}$ that corresponding to $M_z$ as the direction of the current pulse reversed. (c-d) Histogram

of the switching possibility on four different devices, all made from the same stack. The lateral sizes of device 1, 2, 3 and 4 are 100 nm ×400 nm, 80 nm ×320 nm, 50 nm ×300 nm, 40 nm ×280 nm, respectively. e-f. Variation of the threshold amplitude of the switching current as a function of the absolute value of a magnetic field applied along the film normal for current flowing in the (e) +y direction and (f)-y direction.

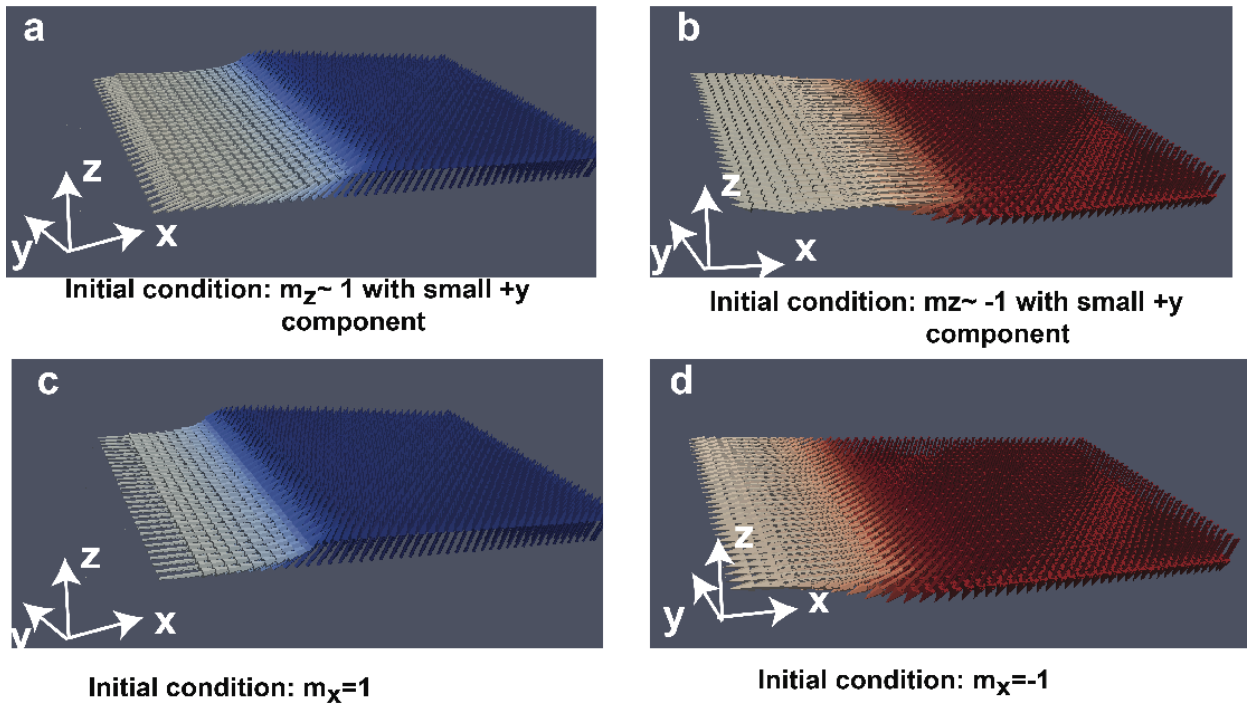

**Figure 5: micromagnetic simulation:** a) Starting from the magnet saturated in +z state with a small +y component, the system is allowed to relax. The final equilibrium state of the magnet is such that the moments in the wedge region point towards +x. Thus average magnetization tilts towards +x. b) Similarly starting from the magnet saturated in -z state, the magnetization tilts towards -x at equilibrium. c) Starting from the magnet in +x , caused by spin accumulation in +x direction due to -y directed current pulse, the magnet evolves to +z state due to the tilt in the anisotropy axis. d) Starting from the magnet in -x , caused by spin accumulation in -x direction due to +y directed current pulse, the magnet evolves to -z state. Blue arrow indicates that the magnetic moment is in +z while the red arrow indicates it is in -z state.